\documentclass[twocolumn,showpacs,preprintnumbers,amsmath,amssymb]{revtex4}
\usepackage{psfig}
\begin{document}

\title{Parametric Spectral Correlations in Disordered and Chaotic Structures}

\author{I. E. Smolyarenko, F. M. Marchetti, and B. D. Simons} 

\affiliation{Cavendish Laboratory, Madingley Road, Cambridge CB3\ OHE, UK}

\date{\today}

\begin{abstract}
We explore the influence of external perturbations on the energy
levels of a Hamiltonian drawn at random from the Gaussian unitary
distribution of Hermitian matrices. By deriving the joint distribution
function of eigenvalues, we obtain the ($n$,$m$)-point parametric
correlation function of the initial and final density of states for
perturbations of arbitrary rank and strength.
A further generalization of these results allows for the incorporation
of short-range spatial correlations in diffusive as well as ballistic
chaotic structures.
\end{abstract}

\pacs{71.23.-k,05.45.Mt,73.21.-b}

\maketitle 


Beginning with early pioneering work on resonances in complex atomic 
nuclei~\cite{porter}, the statistical properties of general chaotic and 
disordered quantum structures have come under intense 
scrutiny~\cite{weidenmueller}. In such systems, it is conjectured
(and, in the case of disordered systems, proven explicitly~\cite{efetov}) 
that the low-energy correlations of eigenvalues (and eigenfunctions) of an 
individual system can be inferred from the {\em ensemble average} properties 
of a random matrix distribution whose members exhibit the same fundamental 
symmetries~\cite{bgs}. For an $N\times N$ matrix Hamiltonian $H$ drawn from, 
say, a Gaussian distribution, the statistical properties of the eigenvalues 
$\{\epsilon_i\}$ are completely specified by the joint distribution 
function 
\begin{equation}
P(\{\epsilon_i\})\propto |\Delta(\{\epsilon_i\})|^\beta 
e^{-(\beta/2)\sum_{i=1}^N\epsilon_i^2},
\label{df}
\end{equation}
where $\Delta(\{\epsilon_i\})=\prod_{i<j}^N\left(\epsilon_i-\epsilon_j
\right)$, and $\beta=1$, $2$ and $4$ for ensembles of orthogonal, unitary
and symplectic symmetry, respectively. From this result, it follows that 
the $n$-point correlator of density of states (DoS) $R_n(\{\Omega_a\})=
\langle \prod_{a=1}^n \nu\left(\Omega_a\right)\rangle$, where 
$\nu\left(\Omega\right)\equiv\bar{\Delta} {\mathrm
tr}\,\delta\left(\Omega-H\right)$, can be expressed as a
determinant~\cite{mehta}
\begin{equation}
\label{eq1}
R_n(\{\Omega_a\})={\mathrm det}_{ab} K_N(\Omega_a,\Omega_b).
\end{equation}
In the Gaussian unitary ensemble (GUE) the kernel 
has the limiting form $\lim_{N\rightarrow\infty}K_N(\Omega_a,\Omega_b)
=\bar{k}\left({\Omega_a-\Omega_b}\right)$, where $\bar{k}\left(\Omega_a-
\Omega_b\right)=k\left(\Omega_a-\Omega_b\right)-\delta\left(\Omega_a-
\Omega_b\right)\theta\left(b-a\right)$ and, defining the average energy 
level spacing $\bar{\Delta}$, 
$k(\Omega)\equiv\sin(\pi\Omega/\bar{\Delta})/(\pi\Omega/\bar{\Delta})$.
While the scaling variable $\bar{\Delta}$ is sensitive to the specific
properties of the system, the function $k(\Omega)$ is universal
provided 
$\Omega \ll {\mathcal E}_c$, where ${\mathcal E}_c^{-1}$ is the time
scale for ergodicity (equivalently, $\Omega/\bar{\Delta}\ll g$, where
$g={\mathcal E}_c/\bar{\Delta}$ is the dimensionless
conductance)~\cite{weidenmueller}.

Often one encounters applications where the interest is in the parametric
dependence of the energy levels of a random Hamiltonian $H'=H+V$ subject
to an extended non-uniform (but slowly varying) perturbation $V$ such
as that imposed by a magnetic field or by an external gate voltage on a
quantum dot (QD). Surprisingly, such correlations exhibit the same
degree of universality: when expressed through the dimensionless
parameter $x^2=X^2 \langle (\partial \epsilon_i/\partial X)^2\rangle$,
where $X$ parameterizes the strength of the external perturbation $V$
(provided the distribution of $H'$ is stationary with respect to $X$),
the statistical properties of the entire random functions
$\epsilon_i(x)$ become universal, dependent only on the fundamental
symmetries of the system~\cite{sa,ls}.

However, there exists a number of important applications where
the perturbation is either local, or rapidly fluctuating, or both.  
One example is presented by the orthogonality catastrophe in a
disordered conductor~\cite{aleiner} (of volume ${\mathcal V}$) where, 
in the coordinate basis, $V({\mathbf r})= v\,{\mathcal V}
\delta^d({\mathbf r}-{\mathbf r}_0)$. Within random matrix theory 
(RMT), this is modeled by a rank $r=1$ perturbation
$V_{ij}=vN\delta_{ij}\delta_{i1}$. The shape-distorting effect of a
gate acting on a ballistic QD can be either global or local, depending
on the relation between the spatial extent of the gate and the
wavelength of the electron $\lambda_F$. Similarly, attempts have been
made to explore the pumping of electrons through a ballistic quantum
dot under the action of two local gates~\cite{pump} which effectively
represent a rank $r=2$ perturbation.  The potential $V({\mathbf
r})=v\,{\mathcal V}[\delta^d({\mathbf r}-{\mathbf r}_1)
-\delta^d({\mathbf r}-{\mathbf r_2})]$ imposed by a bistable impurity
in a disordered metallic host presents another application of a rank
$r=2$ perturbation. A bistable dislocation segment provides an example
of a perturbation of a still higher rank. The freedom to explore
perturbations of high rank is easily afforded in experiments on
microwave resonators with movable scatterers \cite{Barth}, or in STM
devices where the tip does not resolve the Fermi wavelength.

The aim of this letter is to show that, when applied to a member of
the unitary random matrix ensemble, perturbations of arbitrary rank
$r$ can be incorporated into a general scheme which allows analytical
expressions for the joint distribution function and the DoS
correlation function to be determined explicitly. In particular, for a
rank $r$ matrix $V=N\widehat{v}$ whose non-zero eigenvalues form a set
$N\{v_k\}_{k=1}^r$, the ($n,m$)-point parametric correlation function
between the initial and the final DoS $R_{nm}(\{\Omega_a\},\{\Omega_b'\})
=\left\langle \prod_{a=1}^n \nu(\Omega_a)
\,\prod_{b=1}^m \nu'(\Omega'_b)\right\rangle$
can be expressed as the $(n+m)\times (n+m)$ determinant
\begin{multline}
\label{eq2}
\lim_{N\rightarrow\infty}R_{nm}(\{\Omega_a\},\{\Omega_b'\})\\
=\det
\left(\begin{array}{cc}\bar{k}(\Omega_a-\Omega_b) & 
\hat{\mathcal D}^{-1}_{\widehat{v},\Omega_a}
\widetilde{k}(\Omega_a-\Omega'_{b'}) \\
\hat{\mathcal D}_{\widehat{v},\Omega'_{a'}}k(\Omega'_{a'}-\Omega_b) & 
\bar{k}(\Omega'_{a'}-\Omega'_{b'}) 
\end{array} \right).
\end{multline}
Here $\widetilde{k}(\Omega_a-\Omega'_{b'})=k(\Omega_a-\Omega'_{b'})-
\delta(\Omega_a-\Omega'_{b'})$, and 
\begin{equation}
\label{d}
\hat{\mathcal D}_{\widehat{v},\Omega}=
\det\left(\openone-\widehat{v}\frac{d}{d\Omega}\right)\equiv
\prod_{k=1}^r\left(1-v_k \, \frac{d}{d\Omega}\right),
\end{equation} 
while the inverse of $\hat{\mathcal D}_{\widehat{v},\Omega}$ has a
convenient integral representation
%
\[
\hat{\mathcal D}^{-1}_{\widehat{v},\Omega} h(\Omega)=\int \prod_{k=1}^N
\left(d\chi_kd\chi^*_k/\pi\right)\, e^{-\left|\mbox{\boldmath$\chi$}\right|^2} 
h(\Omega+\mbox{\boldmath$\chi$}^\dagger\widehat{v}\mbox{\boldmath$\chi$}).
\]
%
The diagonal blocks of (\ref{eq2}) reproduce the standard correlations
(\ref{eq1}) (see also Ref. \cite{bh}), while, in the particular case
of rank $r=1$, for $n=m=1$ Eq.\ (\ref{eq2}) recovers the result
obtained in Ref.~\cite{aleiner}.  The corresponding joint distribution
function of the combined set of eigenvalues $\{\epsilon_i\}$ of the
matrix $H$ and $\{\epsilon^{\prime}_i\}$ of $H'$ is given by
\begin{equation}
\!\!{\mathcal P}\left(\{\epsilon_i\},\!\{\epsilon^{\prime}_i\}\right) 
\!\propto\! 
{\mathcal P}\left(\{\epsilon_i\}\right)\!
{\Delta(\{\epsilon'_i\})\over\Delta(\{\epsilon_i\})}
{\det}_{ij}\!\left[\hat{\mathcal
D}^{-1}_{\widehat{v},\epsilon_i}\!\delta\left(\epsilon_i
-\epsilon'_j\right)\right]
\label{eq3}
\end{equation}
where ${\mathcal P}(\{\epsilon_i\})$ represents the distribution
function (\ref{df}) of the eigenvalues of the matrix $H$ for
$\beta=2$.

To apply these results to physical systems, it is necessary to effect
a generalization to account for cases where the Hamiltonian dynamics
$H$ is generated by a sum of regular and random components. This leads
to non-vanishing off-diagonal terms in the ensemble average propagator
$\langle G(\epsilon_F)\rangle=\langle(\epsilon_F- H)^{-1}\rangle$. In
disordered metals, the corresponding non-universal terms are known as
Friedel oscillations and reflect the short-ranged ballistic nature of
wave propagation via the Friedel function $f({\mathbf
r})=(\bar{\Delta}{\mathcal V}/\pi) {\mathrm Im}\langle
G(\epsilon_F-i0;0,{\mathbf r})\rangle$. In the presence of an assembly
of rank one perturbations $V({\mathbf r})=\sum_k v_k{\mathcal
V}\delta^d({\mathbf r}- {\mathbf r}_k)$, parametric spectral
correlations are heavily influenced by the spatial arrangement of the
wavefunction around the impurities.  Surprisingly, the correlation
function in this case retains the overall structure of (\ref{eq2}),
and the effect of the Friedel oscillations is incorporated into the
expression for $R_{nm}$ by means of the generalization
\begin{equation}
\label{eq5}
\hat{\mathcal
D}_{\widehat{v},\Omega}\mapsto\hat{\mathcal D}_{{\widehat{f}
\widehat{v}},\Omega}\equiv\det\left(\openone-\widehat{f}\,\widehat{v}\,
\frac{d}{d\Omega}\right),
\end{equation}
where the elements of the $r\times r$ matrix $\widehat{f}$ are prescribed
by the Friedel function $f_{kl}\equiv f({\mathbf r}_k-{\mathbf r}_l)$, 
and $\widehat{v}$ is the diagonal matrix with entries $v_k
\delta_{kl}$. If the points ${\mathbf r}_k$ and ${\mathbf r}_l$ are 
separated by a distance in excess of the mean-free path, $f({\mathbf r}_k
-{\mathbf r}_l)\mapsto \delta_{kl}$, and Eq.\ (\ref{d}) is recovered.

Within RMT, Eq.\ (\ref{eq2}) applies to perturbations of arbitrary
rank and strength. In systems where the Hamiltonian dynamics includes
a regular component (such as diffusive or ballistic conductors) Eqs.\
(\ref{eq2}) and (\ref{eq5}) represent the leading (zeroth) order term
in an expansion in the inverse powers of the dimensionless conductance
$g$. Whether the subleading terms in this expansion are small depends
sensitively on the range and profile of the perturbation. If ${\mathrm
tr}\,(\widehat{f}V)^2> {\mathrm tr}\,(V\Pi V)$, where $\Pi$ is the
diffusion (Perron-Frobenius) propagator in diffusive (ballistic)
systems, the results above apply without qualification \cite{fm,note}.
In the opposite limit, when rescaled by a {\em single} parameter which
incorporates diffusive correlations at length scales in excess of the
mean free path, parametric correlations acquire a universal
form~\cite{sa}, independent of the detailed properties of the
system. The {\em structure} of these correlations can be inferred from
the RMT above: for an extended perturbation of rank $r\sim O(N)$, the
requirement that $\hat{\mathcal D}$ stays finite as
$N\rightarrow\infty$ implies that ${\mathrm tr}\,V^2\sim
O(N^2\bar{\Delta}^2)$. It follows that, for all $n\ge 3$, ${\mathrm
tr}(V/N\bar{\Delta})^{n}
\rightarrow 0$, and $\hat{\mathcal D}_{\widehat{v},\Omega}\rightarrow
\exp[-({\mathrm tr}\,\widehat{v}^2/2)\,d^2/d\Omega^2]$ 
reproducing identically the results of Ref. \cite{sa}.

In the standard Wigner-Dyson RMT, the determinantal structure
(\ref{eq1}) of the many-point correlation functions makes it possible
to obtain analytical results for properties associated with {\em
individual} energy levels, such as level spacing statistics and its
various generalizations \cite{mehta}. Similarly, we anticipate that on
the basis of the $(n,m)$-point parametric correlation function
obtained in this Letter, an analogous theory of `parametric level
spacings' in the GUE can be developed.

To illustrate this point, we consider the evolution of a {\em single}
level under the influence of a rank $r=1$ perturbation. Specifically,
we study the distribution $P(d)=\bar{\Delta}\left\langle
\sum_i\delta\left(\Omega-\epsilon_i\right)
\delta\left[\Omega+d-\epsilon'_i\left(v\right)\right]\right\rangle$ 
of distances $d=\epsilon'_i(v)-\epsilon_i$ between a given level
$\epsilon_i$ and its ``descendant'' $\epsilon'_i(v)$. This quantity is
of independent interest in the context of Coulomb blockade (CB) peak
spacings in disordered QDs. It is known that RMT-based
statistical theory of peak spacings shows poor agreement with
experiment \cite{ABG}. The disagreement is most likely attributable to
the effects of strong Coulomb interaction which manifest themselves as
``spectral scrambling'' \cite{bmm} -- readjustment of single-particle
eigenstates to accommodate the non-uniform spatial distribution of the
charge of an added electron, and the parametric ``gate effect''
\cite{Vallejos} -- distortion of the shape of the QD as the gate
voltage is changed to sweep through a sequence of CB peaks. It would
be interesting to study the latter effect in isolation, for example by
comparing the two sequences of CB peaks corresponding to $v=0$ and
$v\ne 0$. If the ``gate effect'' obeys RMT statistics, the
distribution of differences in peak positions should coincide with
$P(d)$. An $r=1$ perturbation can be simulated, for example, by
adjusting the voltage on a finger-shaped gate whose tip has a spatial
extent smaller than $\lambda_F$.

A particular feature of the $r=1$ perturbation is the identity
$\epsilon_i< \epsilon'_i<\epsilon_{i+1}$~\cite{aleiner}, where the
levels are numbered in increasing order (assuming $v>0$). $P(d)$ thus
coincides with the 
distribution $\tilde{P}(d)$, defined as the probability that an
interval of length $d$ is bounded by a pair
$(\epsilon_i,\epsilon'_j(v))$ at its ends, and does not contain inside
it any levels from either of the two sets. The above identity ensures
$j=i$. The standard method of computing level spacing distributions in
RMT \cite{mehta} affords a straightforward generalization to the
parametric case allowing the distribution $\tilde{P}(d)$ to be
expressed as
\[
\tilde{P}(d)=\det(\openone - K) 
\left[{\mathcal G}_{11}(0,0)
{\mathcal G}_{22}(d,d)-{\mathcal G}_{12}(0,d){\mathcal
G}_{21}(d,0)\right]
\]
(cf. Eq. (5.4.31) of Ref. \cite{mehta}), where the integral kernel
\begin{equation*}
K(\epsilon,\epsilon')=\left(\begin{array}{cc}k\left(\epsilon-\epsilon'\right)
& \hat{\mathcal D}_{\widehat{v},\epsilon}^{-1}\widetilde{k}
\left(\epsilon-\epsilon'\right) \\
\hat{\mathcal D}_{\widehat{v},\epsilon}k\left(\epsilon-\epsilon'\right) &
k\left(\epsilon-\epsilon'\right)
\end{array}\right) 
\end{equation*}
is defined on the segment $[0,d]$ of the real line, and ${\mathcal
G}_{\alpha\beta}\left(\epsilon,\epsilon'\right)$ is the inverse of 
$\openone - K^{-1}$. In Fig. 1 we present 
a set of $\tilde{P}(d)$ curves 
at several different values of $v$. Note that in the limit
$v\rightarrow 0$ the distribution $\tilde{P}_{v\rightarrow
0}(d)=\frac{1}{v}e^{-d/v}$ coincides with the appropriately scaled
level velocity distribution \cite{mucciolo}, while for $v \gtrsim 3$ the
curves $\tilde{P}(d)$ develop a maximum at a finite value of $d$
signifying weak repulsion between the old and the new levels.
\begin{figure}
\centerline{\psfig{file=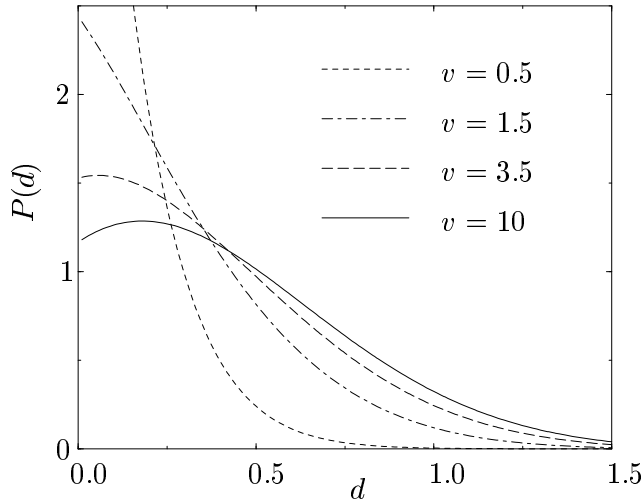,width=3.4in}}
\vspace{0.5cm}
\caption[]{The $P(d)\equiv \tilde{P}(d)$ curves at several 
characteristic values of $v$ for a rank $r=1$ perturbation.}
\label{fig1}
\end{figure}

Let us turn now to the derivation of Eqs.\ (\ref{eq2}), (\ref{d}), and
(\ref{eq5}). To obtain the correlation function $R_{nm}$ one can follow
at least two different routes: By implementing conventional methods of
statistical field theory, $R_{nm}$ can be expressed in the form of a
supersymmetric field integral with an action of non-linear
$\sigma$-model type~\cite{efetov}. Although such an approach has the
advantage of allowing a systematic development of non-universal
corrections in both the disordered and chaotic environments, this
method is largely tailored to the consideration of two-point
correlation functions (i.e. $n=m=1$). Thus, we will follow a
complementary route employing a method based on orthogonal
polynomials~\cite{mehta}. A similar approach was employed in
Ref. \cite{yan} to study the statistics of eigenvalues of matrices
belonging to the GUE under the influence of a finite rank
non-Hermitian perturbation. The latter ensemble serves as a model for
describing resonances in quantum chaotic scattering \cite{yan2}.

The joint distribution of complex $N\times N$ Hermitian matrices $H$
and $H'$ is defined by
\begin{equation}
{\mathcal P}\left(H,H'\right)={\mathcal P}\left(H\right)\delta\left(H'-H-
V\right),
\label{p1}
\end{equation}
where ${\mathcal P}(H)$ represents the Gaussian distribution function
of $H$, and the matrix $\delta$-function is understood as a
product of $N^2$ scalar $\delta$-functions (one per each independent
component of the complex Hermitian matrix $H$). Setting
$H=U\,\widehat{\epsilon}\, U^\dagger$ and
$H'=U'\,\widehat{\epsilon}^{\prime}
\, {U'}^\dagger$
where $\widehat{\epsilon}$ and $\widehat{\epsilon}'$ represent the
diagonal matrices of eigenvalues, the joint distribution function
${\mathcal P}(\{\epsilon_i\},\{\epsilon^{\prime}_i\})$ is obtained
from~(\ref{p1}) by integrating out the `angular' degrees of freedom,
\begin{multline*}
{\mathcal P}(\{\epsilon_i\},\{\epsilon^{\prime}_i\})\\ = \int_{{\rm
U}(N)} \Delta^2(H) d\mu(U)
\int_{{\rm U}(N)} \Delta^2(H') d\mu(U')\, {\mathcal P}\left(H,H'\right),
\end{multline*}
where $d\mu\left(U\right)$ represents the invariant Haar measure on
the unitary group ${\rm U}(N)$, and
$\Delta(H)\equiv\Delta(\{\epsilon_i\})$. To facilitate the angular
integration it is convenient to introduce a Lagrange multiplier in the
form of the Hermitian matrix $\Lambda= U_\Lambda\widehat{\lambda}
U_\Lambda^\dagger$, where $\widehat{\lambda}$ represents the diagonal
matrix of eigenvalues of $\Lambda$, after which the expression for the
joint distribution function can be brought to the form
\begin{widetext}
\begin{equation*}
{\mathcal P}\left(\left\{\epsilon_i\right\},\left\{\epsilon^{\prime}_i
\right\}\right) \propto \int_{{\rm U}(N)} \Delta^2(H) d\mu(U) 
\int_{{\rm U}(N)} \Delta^2(H') d\mu(U')
 \prod_{i=1}^N \int_{-\infty}^\infty d\lambda_i\int_{{\rm
U}(N)} \Delta^2(\Lambda) 
d\mu\left(U_\Lambda\right) 
\,e^{i{\mathrm tr}\,[\Lambda (H+V-H')]-{\mathrm tr}\,H^2}.
\end{equation*}
The overall numerical constant can be restored by demanding the correct
normalization. Integration over $U$ and $U'$ is performed by means
of the Itzykson-Zuber integral~\cite{iz}
\begin{equation*}
\int_{{\rm U}(N)} d\mu\left(U\right) e^{i{\mathrm tr}
\,(AUBU^{\dagger})}\propto
\frac{{\mathrm det}\left[\exp\left(ia_ib_j\right)\right]}
{\Delta(A)\Delta(B)},
\end{equation*}
where $\{a_i\}$, and $\{b_i\}$ represent, respectively, the
eigenvalues of the $N\times N$ Hermitian matrices $A$ and $B$. As a
result,
\begin{equation}
{\mathcal P}\left(\{\epsilon_i\},\{\epsilon^{\prime}_i\}\right) \propto 
{\mathcal P}\left(\{\epsilon_i\}\right) {\Delta(\{\epsilon'_i\})\over
\Delta(\{\epsilon_i\})} \int_{{\rm U}(N)} d\mu\left(U_{\lambda}\right) 
\prod_{i=1}^N \int_{-\infty}^\infty d\lambda_i e^{i{\mathrm
tr}\,(\Lambda V)}
\sum_{{\mathrm P}{\mathrm P}^{\prime}}\left(-1\right)^{{\mathrm P}+
{\mathrm P}'}\exp
\left[i\sum_{i=1}^N\left(\epsilon_{i}\lambda_{{\mathrm P}\left(i\right)}-
\epsilon^{\prime}_{i}\lambda_{{\mathrm P}'\left(i\right)}\right)\right],
\label{10}
\end{equation}
\end{widetext}
\noindent
where the sum over ${\mathrm P}$ runs over all permutations of the
indices $1,...,N$, and $\left(-1\right)^{\mathrm P}$ denotes the
signature of the permutation. However, using the Itzykson-Zuber
formula to undertake the remaining angular integration over
$U_{\Lambda}$ is impractical because of the $N-r$ degenerate (zero)
eigenvalues of the perturbation $V$.  Instead, it is convenient to
express the rank $r$ perturbation in the form $V=N\sum_{k=1}^r v_k
\,{\mathbf a}_k\otimes{\mathbf a}^{\dagger}_k$ where ${\mathbf a}_k$ form 
a set of $r$ complex mutually orthogonal $N$-component vectors of unit
length. Due to the invariance of the measure $\mu(U)$, the integral in
Eq.~(\ref{10}) does not depend on ${\mathbf a}_k$, and therefore its
value does not change upon averaging over them. Although the measure
$d{\mathcal W}$ of the integral over ${\mathbf a}_k$ must formally
enforce the conditions of normalization and mutual orthogonality, in
the large-$N$ limit it can be replaced by the product of the
Porter-Thomas~\cite{pt} distributions of their components:
\begin{equation*}
d{\mathcal W}\left[\left\{{\mathbf a}\right\}\right]=\prod_{k=1}^r
\prod_{i=1}^N \frac{N da_{ki}da^*_{ki}}{\pi} e^{-N\left|a_{ki}\right|^2}.
\end{equation*}
Absorbing the rotations of $\Lambda$ into the invariant measure 
$d{\mathcal W}$, and integrating out the eigenvalues of $\Lambda$, one
obtains
\begin{multline}
{\mathcal P}\left(\{\epsilon_i\},\{\epsilon^{\prime}_i\}\right) \propto
{\mathcal P}\left(\{\epsilon_i\}\right) {\Delta(\{\epsilon'_i\})\over
\Delta(\{\epsilon_i\})}\sum_{{\mathrm P}{\mathrm P}'}(-1)^{{\mathrm P}
+{\mathrm P}'} \\
\times\int d{\mathcal W}\left[\left\{{\mathbf a}\right\}\right]
\prod_{i=1}^N\delta\left(\epsilon'_{{\mathrm P}'(i)}-
\epsilon_{{\mathrm P}(i)}-N\sum_{k=1}^r v_k \left|a_{ki}\right|^2\right).
\label{eq12}
\end{multline}
Then, introducing new variables $\chi_{ki}=\sqrt{N}a_{ki}$, the integral
over $\chi_{ki}$ is identified as the integral representation of
$\prod_{i=1}^N\hat{\mathcal D}^{-1}_{\widehat{v},\epsilon_i}$, thus
establishing Eq.~(\ref{eq3}).

To obtain the correlation function $R_{nm}$ the joint distribution
function~(\ref{eq12}) is multiplied by
%
$\prod_{a=1}^n \sum_{i_a=1}^N\delta(\Omega_a-\epsilon_{i_a})\,\prod_{b=1}^m
\sum_{j_b=1}^N\delta(\Omega'_b-\epsilon'_{j_b})$
%
and then integrated over all eigenvalues $\{\epsilon_i\}$ and
$\{\epsilon_i'\}$. To perform the integrals over $\{\epsilon_i\}$ we
utilize the following identity: $\prod_{i=1}^N\hat{\mathcal
D}^{-1}_{\widehat{v},\epsilon_i}\Delta\left(\{\epsilon_i\}\right)=
\Delta\left(\{\epsilon_i\}\right)$. Upon integrating by parts over all 
$\epsilon_i$ except the $n$ fixed energies $\epsilon_{i_a}$, and
integrating over all $\epsilon_j'$, the correlation function can be
recast as~\cite{note2}
\begin{multline}
\label{rnm}
R_{nm}(\{\Omega_a\},\{\Omega_b'\})
\propto 
\sum_{{i_1...i_n}\atop {j_1...j_m}}\int\prod_{i=1}^Ne^{-\epsilon_i^2}
d\epsilon_i \\
\times\left[\left(\prod_{a=1}^n \delta\left(\epsilon_{i_a}-
\Omega_a\right)\hat{\mathcal D}_{\widehat{v},\epsilon_{i_a}}\right) \Delta
\left(\{\epsilon_i\}\right)\right]\\
\times\left[\left(\prod_{a=1}^n
\hat{\mathcal D}^{-1}_{\widehat{v},\epsilon_{i_a}}\right)
\Delta\left(\left\{\epsilon_i\right\}\right)\prod_{b=1}^m
\delta\left(\epsilon_{j_b}-\Omega_b'\right)\right].
\end{multline}
After some algebra, Eq.\ (\ref{eq2}) is straightforwardly established
following the standard procedure of expanding the Vandermonde
determinants in terms of Hermite polynomials and using their
orthogonality properties.

Finally, to establish the form of $\hat{\mathcal D}$ in the case when
Friedel correlations are present, we note that $V_{ij}$ can be written
as $\sum_{lp}U_{il}[\sum_{k=1}^r v_k \,{\mathbf \psi}^{(l)}_k {\mathbf
\psi}^{(p)\dagger}_k] U_{pj}^{\dagger}$, where $\psi^{(l)}_k$ is the
$k$-th component of the $l$-th eigenfunction of $H$. To the
leading order in $1/g$, the distribution of $\psi^{(l)}_k$ is
uncorrelated with $\left\{\epsilon_i\right\}$, and is given by Berry's
conjecture \cite{Berry,Mirlin}. In particular, in the unitary
ensemble,
\begin{equation*}
d{\mathcal W}\left[\left\{\psi\right\}\right]\propto \prod_{kl}
d\psi_k^{(l)*}d\psi_k^{(l)}
e^{-\sum_l\psi^{(l)*}_k\left(\widehat{f}^{-1}\right)_{kk'}\psi^{(l)}_{k'}}.
\end{equation*}
Absorbing $U$ into $U_\lambda$ and reversing the order of integration
over $d{\mathcal W}$ and $U'$, we arrive at Eqs.\ (\ref{eq3}),
(\ref{eq5}).

To summarize, we have demonstrated that the $(n,m)$-point parametric
correlation function of the DoS in GUE possesses a determinantal
structure for perturbations of arbitrary rank. Beyond GUE, a
generalization of this determinantal structure is shown to accommodate
spatial correlations induced by ballistic wave propagation at length
scales shorter than the mean free path.





\begin{thebibliography}{}

\bibitem{porter} For a review, see {\em Statistical theories of spectra: 
fluctuations}, ed. C. E. Porter (Academic Press, New York, 1965).

\bibitem{weidenmueller} For a review, see e.g., T. Guhr, A. 
M\"uller-Groeling, and H. A. Weidenm\"uller, Phys. Rep. {\bf 299}, 189 
(1998).

\bibitem{efetov} K. B. Efetov, Adv. Phys. {\bf 32}, 53 (1983).

\bibitem{bgs}O.~Bohigas, M.~J.~Giannoni, and C.~Schmit,
Phys. Rev. Lett. {\bf 52}, 1 (1984); J. Physique Lett. {\bf 45}, L1615
(1984).

\bibitem{mehta} M. L. Mehta, {\em Random Matrices} (Academic Press,
New York, 1991).

\bibitem{sa} For a review, see B. L. Altshuler and B. D. Simons, in 
``Mesoscopic Quantum Physics'', eds. E Akkermans {\em et al.}, Les
Houches, (North-Holland, Amsterdam, 1994).

\bibitem{ls}See P.~Leboeuf and M.~Sieber, Phys. Rev. E {\bf 60}, 3969
(1999) for the discussion of ensembles with non-stationary parametric
dependence.

\bibitem{aleiner}I.~L.~Aleiner and K.~A.~Matveev,
Phys. Rev. Lett. {\bf 80}, 814 (1998).

\bibitem{pump}M. Switkes {\em et al.}, Science {\bf 283}, 1905 (1999).

\bibitem{Barth} M.~Barth, U.~Kuhl, and H.-J.~St\"{o}ckmann,
Phys. Rev. Lett. {\bf 82}, 2026 (1999).

\bibitem{bh}E.~Br\'{e}zin and S.~Hikami, Phys. Rev. E {\bf 56}, 264
(1997).


\bibitem{fm}F.~M.~Marchetti, I.~E.~Smolyarenko, and B.~D.~Simons, in 
preparation.

\bibitem{note}The l.h.s of this inequality is dominated by rapidly 
oscillating components of $V$, while the r.h.s is strongly weighted
towards its slow modes.

\bibitem{ABG}For recent reviews, see I.~L.~Aleiner, P.~W.~Brouwer, and
L.~I.~Glazman, Phys. Rep. {\bf 358}, 309 (2002); Y.~Alhassid,
Rev. Mod. Phys. {\bf 72}, 895 (2000).

\bibitem{bmm}Ya.~M.~Blanter, A.~D.~Mirlin, and B.~A.~Muzykantskii,
Phys. Rev. Lett. {\bf 78}, 2449 (1997).

\bibitem{Vallejos}R.~O.~Vallejos, C.~H.~Lewenkopf, and E.~R.~Mucciolo,
Phys. Rev. Lett. {\bf 81}, 677 (1998).

\bibitem{mucciolo}E.~R.~Mucciolo, V.~N.~Prigodin, and B.~L.~Altshuler,
Phys. Rev. B {\bf 51}, 1714 (1995).

\bibitem{yan}Y.~V.~Fyodorov and B.~A.~Khoruzhenko,
Phys. Rev. Lett. {\bf 83}, 65 (1999).

\bibitem{yan2}Y. V. Fyodorov and H.-J. Sommers, J. Math. Phys. {\bf
38}, 1918 (1997). 

\bibitem{iz}C. Itzykson and J. -B. Zuber, J. Math. Phys. {\bf 21}, 411
(1980). 

\bibitem{pt} C. E. Porter and R. G. Thomas, Phys. Rev. {\bf 104}, 483
(1956).

\bibitem{note2}The action of the differential operator on the
confining potential has been neglected; it produces $O(1/N)$
corrections as long as all energies are close to the center of the
Wigner semicircle. Note also that the operator $\hat{\mathcal D}^{-1}$
in the second set of square brackets in Eq.\ (\ref{rnm}) acts on both
the Vandermonde determinant and the $\delta$-functions, which accounts
for the $\delta$-function terms in the upper right elements of the
matrix in Eq.\ (\ref{eq2}).

\bibitem{Berry}M. V. Berry, J. Phys. A {\bf 10}, 2083 (1977);
S.~Hortikar and M.~Srednicki, Phys. Rev. Lett. {\bf 80}, 1646 (1998).

\bibitem{Mirlin}I.~V.~Gornyi and A.~D.~Mirlin, Phys. Rev. E {\bf 65},
025202(R) (2002). 


\end{thebibliography}
\end{document}